\PassOptionsToPackage{dvipsnames}{xcolor}
\documentclass[longbibliography,nofootinbib,aps,prx,superscriptaddress,twocolumn]{revtex4-2}

\usepackage{amsmath,amsfonts,amssymb,amsthm,dcolumn,bm,qcircuit,appendix,mathtools,thmtools,thm-restate,algorithm,algpseudocode,mathrsfs,pgfplots,soul,lipsum,graphicx,braket,enumitem,caption,subcaption,ragged2e}

\usepackage[
  labelfont=bf,
  labelsep=period,
  justification=justified,
  format=plain,
  singlelinecheck=false
]{caption}

\usepackage{pgfplots}
\usepgfplotslibrary{patchplots}
\usetikzlibrary{3d, decorations.pathmorphing}
\usepackage{tikz,tkz-euclide,tikz-3dplot}
\pgfplotsset{compat=newest}
\usepgfplotslibrary{colormaps}
\usepgfplotslibrary{patchplots}
\usepackage{booktabs}

\usetikzlibrary{matrix,fit,backgrounds,3d,arrows, decorations.pathreplacing,shapes.geometric,3d, calc}

\def\BibTeX{{\rm B\kern-.05em{\sc i\kern-.025em b}\kern-.08em
    T\kern-.1667em\lower.7ex\hbox{E}\kern-.125emX}}

\newtheorem{theorem}{Theorem}
\newtheorem{lemma}{Lemma}

\newtheorem{remark}{Remark}
\newtheorem{definition}{Definition}
\newtheorem{method}{Algorithm}

\numberwithin{proposition}{section}
\numberwithin{theorem}{section}
\numberwithin{lemma}{section}
\numberwithin{corollary}{section}
\numberwithin{remark}{section}
\numberwithin{definition}{section}
\numberwithin{equation}{section}

\DeclareMathOperator{\Tr}{Tr}

\newcommand{\Ibb}{\mathbb{I}}

\usepackage{xcolor}
\usepackage{hyperref}

\hypersetup{
    colorlinks=true,
    linkcolor=MidnightBlue,
    filecolor=magenta,
    urlcolor=teal,
    pdfpagemode=FullScreen,
    citecolor=OliveGreen
}

\begin{document}
\title{Quantum Algorithms for Computing Maximal Quantum $f$-divergence and Kubo-Ando means}
\author{Trung Hoa Dinh}
\email{{thdinh@troy.edu}}
\affiliation{Department of Mathematics and Statistics, Troy University, \\ Alabama 36082, USA}

\author{Nhat A. Nghiem}
\email{{nhatanh.nghiemvu@stonybrook.edu}}
\affiliation{Department of Physics and Astronomy, State University of New York at Stony Brook, \\ Stony Brook, NY 11794-3800, USA}
\affiliation{C. N. Yang Institute for Theoretical Physics, State University of New York at Stony Brook, \\ Stony Brook, NY 11794-3840, USA}

\begin{abstract}
The development of quantum computation has resulted in many quantum algorithms for a wide array of tasks. Recently, there is a growing interest in using quantum computing techniques to estimate or compute quantum information-theoretic quantities such as Renyi entropy, Von Neumann entropy, matrix means, etc. Motivated by these results, we present quantum algorithms for computing the maximal quantum $f$-divergences and the operator-theoretic matrix Kubo--Ando means. Both of them involve Renyi entropies, matrix means as special cases, thus implying the universality of our framework. 
\end{abstract}

\maketitle

\section{Introduction}
Quantum computation has advanced a long path, with many quantum algorithms discovered capable of solving many challenging computational hurdles. Notable examples include Grover's search algorithm \cite{grover1996fast}, Shor's factorization algorithm \cite{shor1999polynomial, kitaev1995quantum}, black-box computing \cite{deutsch1985quantum,deutsch1992rapid}, quantum simulation algorithms \cite{berry2007efficient,berry2017quantum,berry2015simulating,berry2012black,berry2015hamiltonian, childs2017quantum,childs2018toward, lloyd1996universal}, quantum machine learning algorithms \cite{lloyd2013quantum,lloyd2014quantum, schuld2018supervised,schuld2019evaluating,schuld2019machine, schuld2019quantum, mitarai2018quantum}, etc. 

Recently, some efforts have been devoted to developing quantum algorithms for computing physical quantities, including Renyi entropy \cite{acharya2020estimating, subramanian2021quantum, wang2023quantum}, Von Neumann entropy \cite{nghiem2025estimation, li2018quantum}. These quantities are fundamental in quantum information theory \cite{wilde2013quantum}, exhibiting both theoretical and practical impact. In a parallel aspect, matrix means of positive definite (PD) matrices are also fundamental in quantum information geometry, statistics, and optimization, with applications in quantum state analysis and machine learning \cite{bhatia2009positive, petz2007quantum}. Computing means such as the geometric mean, log--Euclidean mean, Karcher mean, and Bures--Wasserstein barycenter require evaluating nonlinear matrix functions (e.g., \( x^{1/2} \), \( \log x \), \( \exp x \)). 
The prior work \cite{liu2025quantum} introduced quantum subroutines for the standard and weighted geometric means, embedding them into unitary operators using block-encoding and addressing nonlinear matrix equations like the Riccati equation. 

This paper builds on the foundation of the aforementioned works, as we extend the curiosity to more general quantities. They include \textit{maximal quantum $f$-divergences}, which involve Renyi/Von Neumann entropy as special cases, and Kubo-Ando means, which involve both standard and weighted geometric means as special cases. As such, our soon-to-be-outlined algorithm is universal, and can be adjusted with little effort to produce the specific entropy, or matrix-mean of interest. 

Our work is organized as follows. Section \ref{sec: overview} provides an overview of our work, including the key techniques we will use, followed by a statement of our main results. Section \ref{sec: quantumalgorithmfivergence} contains our quantum algorithm for $f$-divergence. Section \ref{sec: quantumalgorithm} is devoted to the quantum algorithm for computing Kubo–Ando means, which also encompasses the geometric means studied in \cite{liu2025quantum}. Finally, we provide a conclusion in Section \ref{sec: conclusion}.

\section{Overview of objectives, techniques and main results}
\label{sec: overview}
The first key objective of our work is, given appropriate access to density matrices $\sigma, \rho$, to estimate the so-called maximal quantum $f$-divergence \cite{petz1998contraction}: 
\begin{align}
    D_f^{\max}(\rho\|\sigma)
= \operatorname{Tr}\!\left[\sigma^{1/2}\, f\!\big(\sigma^{-1/2}\rho\sigma^{-1/2}\big)\, \sigma^{1/2}\right],
\end{align}
where $f(\cdot)$ belongs to a class of functions called operator-convex functions. It can be seen that the above quantity depends explicitly on the choice of $f(\cdot)$, e.g., $f(x) = x\log x, x^{\alpha}, (x-1)^2$, etc. (as will be shown later, each specific form of $f(\cdot)$ corresponds to the type of known quantum information-theoretic notion). However, it is a known result that any general operator-convex function admits polynomial approximation (with appropriate coefficients). Therefore, we will construct quantum algorithms according to the given choice of $f(\cdot)$, before generalizing them to arbitrary $f(\cdot)$. 

The second objective of our work is to construct a quantum algorithm for Kubo-Ando means $\sigma_f$. Roughly speaking, an operator mean $\sigma_f$ is a binary operation that takes two operators and produces another operator. If $\sigma_f$ satisfies certain properties, then it is called Kubo-Ando mean, e.g., see Part 3 of \ref{sec: preliminaries}. In our work, we denote $A \sigma_f B$ as the outcome of $\sigma_f$ acting on $A,B$. Our key objective then is to compute $A\sigma_f B$, provided appropriate access to $A$ and $B$, as will be discussed below.

At the core of our algorithm in this work is the concept of block-encoding and the quantum singular value transformation framework (QSVT) \cite{gilyen2019quantum}. A more formal and detailed description of the block-encoding/QSVT framework is left in the Appendix \ref{sec: preliminaries}. Here, we briefly revise a few necessary concepts and notations. A unitary $U$ is said to be an exact block-encoding of a matrix $A$ (with operator norm $||A||_{\rm o}\leq 1$, or equivalently, its eigenvalues have magnitude $\leq 1$) if $A = ( \bra{\bf 0}\otimes \Ibb)  U (\Ibb \otimes \ket{0})$. If $ ( \bra{\bf 0}\otimes \Ibb)  U (\Ibb \otimes \ket{0}) = \tilde{A}$ and $||\tilde{A}-A||_{\rm o} \leq \epsilon$, then $U$ is said to be an $\epsilon$-approximated block-encoding of $A$. There exist a known procedure, that can take different block-encoded operators and form another block-encoded operator, which can be product, summation, tensor product, etc, of the original ones. In particular, a fundamental result of QSVT \cite{gilyen2019quantum} is that, given a unitary $U$ that block-encodes $A$, then there the unitary block-encoding of $f(A)$ can be constructed efficiently, where $f(\cdot)$ is some computable functions. 

Regarding our objectives mentioned earlier, in order to estimate the quantum $f$-divergence and Kubo-Ando means, we need a form of access to $\sigma,\rho, A,B$. In order to apply the block-encoding/QSVT framework, it is apparent that we need to be able to block-encode these operators. To this end, we mention the following results, which we assume to have in our work.
\begin{lemma}[Block-encoding density operator \cite{gilyen2019quantum, nghiem2025new1}]
\label{lemma: blockencodingdensityoperator}
The block-encoding of density matrix $\rho \in \mathbb{C}^{N \times N}$ of interested can be obtained via the following ways.
\begin{itemize}
    \item (Purification access \cite{gilyen2019quantum}) Let $U$ be the unitary purification of $\rho$, i.e., $\rho = \Tr_{\rm ancilla } U \ket{\bf 0} \bra{\bf 0} U^\dagger$  where $\Tr_{\rm ancilla }$ refers to tracing out the ancilla system (those extra qubits other than those corresponds to $\rho$). If $U$ is provided, then the (exact) block-encoding of $\rho$ can be obtained using $U_{\rho}, U_{\rho}^\dagger$ once each and another $\log N$ gates. 
    \item (Sample access \cite{nghiem2025new1}) If instead multiple copies of $\rho$ are given, then there exists a quantum procedure that return a $\epsilon$-approximated unitary block-encoding of $\frac{\pi\rho}{4}$. The procedure uses $\mathcal{O}\left( \frac{1}{\epsilon^2}\right) $ copies of $\rho$ and a $\log N$-qubits quantum circuit of depth $\mathcal{O}\left(  \log N \frac{1}{\epsilon^2} \log \frac{1}{\epsilon} \right) $. 
\end{itemize}
\end{lemma}

\begin{lemma}[Block-encoding accessible matrices]
    \label{lemma: blockencodingmatrix}
    The block-encoding of the matrix $A \in \mathbb{C}^{N \times N}$ (assuming without loss of generalization that its operator norm $||A||_{o} \leq 1$) can be obtained via the following ways.
    \begin{itemize}
        \item (Sparse-accessible matrix \cite{gilyen2019quantum}) Given two oracles, or black-box circuits $O_r$ and $O_A$, which act as follows:
\begin{align*}
    O_r \ket{i}\ket{k} &= \ket{i} \ket{r_{ik}}, \\
    O_A \ket{i} \ket{j} \ket{\bf 0} &= \ket{i}\ket{j} \ket{A_{ij}},
\end{align*}
where $1\leq i,j \leq N$ and $k \leq s$ ($s$ is the sparsity -- the maximum number of nonzero entries in each row or column). Then the block-encoding of $\frac{A}{s}$ can be constructed using a quantum circuit of complexity $\mathcal{O}\left(  \log N + \log^{2.5}\frac{1}{\epsilon}\right) $. 
\item (Classically known entries \cite{nghiem2025refined}) If instead, the entries of $A$ are classically, explicitly known, then there is a quantum algorithm using a circuit of depth $\mathcal{O}\left(||A||_F \log (N) \kappa \log^2 \frac{\kappa}{\epsilon} \right)$ which is a $\epsilon$-approximated block-encoding of $A$. Let $s_A$ denotes the total number of nonzero entries of $A$. The total number of qubits for this circuit can be $\mathcal{O}\left( \log N + s_A \right) $, $\mathcal{O}\left( \log N + \log s_A \right) $, or $ \mathcal{O}\left( \log N \right)$ depending on whether $A$ admits a certain decomposable structure. 
    \end{itemize}
\end{lemma}

\begin{lemma}[Approximation of $\log x$]
\label{lemma: approximationoflog}
For $ 0 \leq x \leq 1$, the function $\log x$ can be approximated via two ways:
\begin{itemize}
    \item (Approach 1 \cite{gilyen2019quantum}) Given $\beta \in (0,1]$ and $\epsilon \in (0,1/2]$. Then on the interval $[\beta, 1]$, there exists a polynomial $P(x)$ with $-1 \leq |P(x)| \leq 1$ such that:
\begin{align*}
    \Big| P(x) - \frac{\log x}{2 \log \beta }\Big| \leq \epsilon. 
\end{align*}
Moreover, the degree of $P(x)$ is $\deg (P(x)) = \mathcal{O}\Big( \frac{1}{\beta} \log( \frac{1}{\epsilon} )  \Big).$ Therefore, $x\log x$ can also be approximated by $x P(x)$, which has degree $\mathcal{O}\Big( \frac{1}{\beta} \log( \frac{1}{\epsilon} )  \Big) $.
    \item (Approach 2) Given $\beta \in (0,1]$ and $\epsilon \in (0,1/2]$ as above. The so-called Stieltjes representation of logarithmical function is:
\[
\log x \;=\; \int_{0}^{\infty}\!\!\Big(\tfrac{1}{1+t}-\tfrac{1}{x+t}\Big)\,dt
\]
The function above can be approximated by a positive quadrature:
$$\log x \approx \sum_{k=1}^m \omega_k\big(\alpha_k - (x+\tau_k)^{-1}\big) $$
with $\omega_k>0$, $\tau_k\ge0$, which results in the following approximation:
$$ x \log x \approx  x \sum_{k=1}^m \omega_k\big(\alpha_k - (x+\tau_k)^{-1}\big)$$
In particular, on the inverval $(\beta,1)$, for $m=\Theta\!\big(\log (\frac{1}{\beta} ) \cdot \log \frac{1}{\epsilon}\big)$, it holds that 
$$ || x\log x-  x \sum_{k=1}^m \omega_k\big(\alpha_k - (x+\tau_k)^{-1}\big) || \leq  \epsilon$$
and therefore:
$$|| x \frac{\log x}{\log \beta }-  x \sum_{k=1}^m \frac{1}{\log \beta} \omega_k\big(\alpha_k - (x+\tau_k)^{-1}\big) || \leq  \frac{\epsilon}{|\log \beta|} \leq \epsilon $$
In our subsequent discussion, by a slight abuse of notation, we set $\frac{1}{\log \beta} \omega_k \equiv \omega_k $ and use $x \sum_{k=1}^m \frac{1}{\log \beta} \omega_k\big(\alpha_k - (x+\tau_k)^{-1}\big)  $ as a proper approximation of $ x \frac{\log x}{\log \beta }$. 
\end{itemize}
\end{lemma}

With the above tools, we will show subsequently that the following results are obtained:
\begin{theorem}[Estimating quantum $f$-divergence]
    Let $\rho, \sigma \in \mathbb{C}^{N\times N}$ be two density matrices of interest. Let $\kappa_\rho,\kappa_\sigma$ be the condition number of them. Let $\gamma \equiv \frac{1}{4\kappa_\sigma} \sigma^{-1/2} \rho \sigma^{-1/2}$ and $\kappa_\gamma$ be its condition number.
    \begin{itemize}
        \item If the unitary purification of $\rho,\sigma$ is provided (with circuit complexity $T$), then the quantum $f$-divergence $ D_f^{\max}(\rho\|\sigma)$ (for arbitrary $f$ being operator-convex function) can be estimated to an additive accuracy $\epsilon$ with circuit complexity 
       $$  \mathcal{O} \left(   (2T + \log N)  \kappa^2_{\sigma} \kappa^2_\gamma\log( \kappa_{\sigma})  \log^3(\kappa_\gamma) \frac{1}{\epsilon}\log^4\frac{1}{\epsilon}\right).$$
        \item If instead, independent copies of $\rho,\sigma$ are provided, then the normalized quantum $f$-divergence $D_f^{\max}(\rho\|\sigma)/N$ can be estimated by repeating a quantum circuit of complexity 
      $$ \mathcal{O}\left(  \log (N)  \kappa_\sigma \kappa^2_\gamma \log^2( \kappa_\gamma) \frac{1}{\epsilon^2} \log^5  \frac{1}{\epsilon}  \right)$$
        The number of repetitions is 
        $$ \mathcal{O}\left( \frac{\kappa_{\sigma}^2 \log^2( 4\kappa_{\sigma})  \log^2 \kappa_{\gamma}}{\epsilon^2} \right)  $$
    \end{itemize}
\end{theorem}

\begin{theorem}[Kubo-Ando means ]

Let $A,B $ be two positive matrices with eigenvalues in $[\delta,1]$ (for $\delta \in (0,1)$). Then the Kubo-Ando means $\sigma_f$ can be constructed using a quantum circuit of complexity:
    $$\mathcal{O}\! \left(  \big(  C_A \frac{1}{\delta}\log \frac{1}{\epsilon} +C_B\big)\frac{1}{\delta^2}\,\log^2 (\frac{1}{\varepsilon} \log\frac{1}{\epsilon})\right)$$
    where $C_A,C_B$ are the complexity for block-encoding $A,B$ and $\epsilon$ is the error tolerance. 
\end{theorem}
In the following, we will proceed to describe our main quantum algorithms. 
\section{Quantum algorithms for maximal quantum $f$-divergence}
\label{sec: quantumalgorithmfivergence}
Let \(f:(0,\infty)\to\mathbb{R}\) be an \emph{operator convex} function satisfying \(f(1)=0\).
Recall that \(f\) is operator convex if
\begin{align}
\begin{split}
    f(tA+(1-t)B) \le t f(A) + (1-t) f(B) \\
    (A,B>0,\ t\in[0,1])
\end{split}
\end{align}
This property guaranties that the following map
\begin{align}
    (\rho,\sigma)\ \mapsto\ \sigma^{1/2}\, f\!\left(\sigma^{-1/2}\rho\sigma^{-1/2}\right)\, \sigma^{1/2}
\end{align}
is jointly convex in \((\rho,\sigma)\), and thus suitable for defining monotone quantum divergences under completely positive trace-preserving (CPTP) maps.

Let \(\rho,\sigma>0\) be density operators on a finite-dimensional Hilbert space \(\mathcal{H}\).
The \emph{maximal quantum \(f\)-divergence} associated with the operator convex function \(f\) is defined by
\begin{equation}\label{eq:Dfmax}
D_f^{\max}(\rho\|\sigma)
:= \operatorname{Tr}\!\left[\sigma^{1/2}\, f\!\big(L_\rho R_{\sigma^{-1}}\big)\,\sigma^{1/2}\right],
\end{equation}
where \(L_\rho(X)=\rho X\) and \(R_{\sigma^{-1}}(X)=X\sigma^{-1}\) denote the left and right multiplication operators on \(B(\mathcal{H})\).

In the finite-dimensional case, this can equivalently be written as
\begin{equation}\label{eq:DfmaxAlt}
D_f^{\max}(\rho\|\sigma)
= \operatorname{Tr}\!\left[\sigma^{1/2}\, f\!\big(\sigma^{-1/2}\rho\sigma^{-1/2}\big)\, \sigma^{1/2}\right].
\end{equation}
When \(\rho\) and \(\sigma\) commute, \eqref{eq:DfmaxAlt} reduces to the classical \(f\)-divergence:
\begin{align}
    D_f^{\max}(\rho\|\sigma)=\sum_i \sigma_i f\!\left(\frac{\rho_i}{\sigma_i}\right)
\end{align}
where \(\rho_i,\sigma_i\) are the eigenvalues of \(\rho,\sigma\).

The functional \(D_f^{\max}\) satisfies:

\begin{itemize}
    \item \textbf{Joint convexity:} For any probability weights \(t_i\),
    \[
    D_f^{\max}\!\left(\sum_i t_i\rho_i \Big\| \sum_i t_i\sigma_i\right)
    \le \sum_i t_i D_f^{\max}(\rho_i\|\sigma_i).
    \]
    \item \textbf{Monotonicity:} For any completely positive trace-preserving (CPTP) map \(\Phi\),
    \[
    D_f^{\max}\!\big(\Phi(\rho)\big\|\Phi(\sigma)\big)
    \le D_f^{\max}(\rho\|\sigma).
    \]
    \item \textbf{Maximality:} Among all monotone quantum \(f\)-divergences satisfying the above properties, \(D_f^{\max}\) is the largest one:
    \[
    D_f^{\text{quantum}}(\rho\|\sigma)\ \le\ D_f^{\max}(\rho\|\sigma),
    \]
    for every other admissible monotone divergence corresponding to the same operator convex \(f\).
\end{itemize}

Hence, operator convexity of \(f\) is the essential analytic condition ensuring that \(D_f^{\max}\) is well-defined, jointly convex, and monotone under CPTP maps.

\textbf{ Examples of \(f\)-functions and corresponding divergences:}

\begin{widetext}
    \begin{center}
\begin{tabular}{lll}
\hline
\(f(x)\) & Operator convex? & Divergence \(D_f^{\max}(\rho\|\sigma)\) \\
\hline
\(x\log x\) & Yes & Umegaki relative entropy \\
\((x-1)^2\) & Yes & Quantum $\chi^2$-divergence \\
\(x^\alpha\) for $0<\alpha<1$ & Yes & Petz--R\'enyi type divergence \\
\(x\log x - x + 1\) & Yes & Kullback--Leibler form \\
\hline
\end{tabular}
\label{tab: entropy}
\end{center}

\end{widetext}

In the following, we proceed to describe quantum algorithm for estimating $D_{x\log x}(\rho\Vert\sigma)$ and more generally, the operator-convex case.

\subsection{Case $f(x)=x\log x$ }

\begin{center}
    \textbf{Procedure}
\end{center}
Recall the $f$-divergence of interest is:
\[
D_f(\rho\Vert\sigma) \;=\; \text{Tr}\!\Big[ \sigma^{1/2}\, f\!\big(\sigma^{-1/2}\rho\,\sigma^{-1/2}\big)\, \sigma^{1/2} \Big],
\qquad \rho,\sigma>0.
\]
For $f(x)=x\log x$, we have that:
\begin{align}
    f\!\big(\sigma^{-1/2}\rho\,\sigma^{-1/2}\big) = \left( \sigma^{-1/2}\rho\,\sigma^{-1/2}\right) \log \left( \sigma^{-1/2}\rho\,\sigma^{-1/2}\right)
\end{align}

\begin{method}
\label{algo: algorithm1}
   Assuming condition numbers $\kappa_{\sigma}, \kappa_{\rho}$ of $\sigma, \rho$, respectively. The quantum procedure to estimate $D_{x\log x} (\rho\Vert\sigma)$ obeys the following steps:
\end{method}
\vspace{2mm}
\noindent
\textbf{Step 1: Block-encoding of $\rho$ and $\sigma$.} This can be done via Lemma \ref{lemma: blockencodingdensityoperator}. 

\vspace{2mm}
\noindent
\textbf{Step 2: Forming the block-encoding of $\sim\sigma^{-1/2}\rho\,\sigma^{-1/2}$.} This can be done by first using Lemma \ref{lemma: negative} with $c=-1/2$ to transform the block-encoded operator $\sigma$ into a $\epsilon$-approximated $ \frac{\sigma^{-1/2}}{2\kappa^{1/2}_{\sigma}}$. Then we use Lemma \ref{lemma: product} to form the $\epsilon$-approximated block-encoding of $ \frac{1}{4 \kappa_{\sigma}} \sigma^{-1/2}\rho\,\sigma^{-1/2} $ . For simplicity we define $\gamma \equiv \frac{1}{4 \kappa_{\sigma}} \sigma^{-1/2}\rho\,\sigma^{-1/2} $ and denote $\kappa_{\gamma}$ as the condition number of $\gamma$.

\vspace{2mm}
\noindent
\textbf{Step 3: Obtaining the block-encoding of $ \sim\frac{\gamma \log \gamma}{ \log \kappa_\gamma}  $.} This can be done by combining Lemma \ref{lemma: qsvt} and Lemma \ref{lemma: approximationoflog}. As Lemma \ref{lemma: approximationoflog} contains two approaches to approximate the logarithmic of the function, there are two routes to obtain the block-encoding of $ \gamma \log \gamma $.
\begin{itemize}
    \item (Route 1: direct QSVT-- corresponding to Approach 1 in Lemma \ref{lemma: approximationoflog}) Applying Lemma \ref{lemma: qsvt} directly with the polynomial $xP(x)$ indicated in the \textit{Approach 1} of Lemma \ref{lemma: approximationoflog}. The result is the $\mathcal{O}(\epsilon)$-approximated block-encoding of:
    $$ \frac{\gamma \log \gamma}{2 \log \kappa_\gamma}  $$
    \item (Route 2: via Stieltjes representation-- corresponding to Approach 2 in Lemma \ref{lemma: approximationoflog}) As indicated in the second approach of Lemma \ref{lemma: approximationoflog}, the operator $\gamma \log \gamma $ is can be approximated as:
    \begin{align}
        (\sum_{k=1}^m \omega_k \alpha_k )\gamma  - \gamma \sum_{k=1}^m \omega_k (\gamma + \tau_k \Ibb)^{-1}
    \end{align}
    We recall a property (see \ref{remark: propertyofblockencoding} in the Appendix \ref{sec: preliminaries}) that the block-encoding of identity matrix $\Ibb$ can be simply obtained. Then we can use Lemma \ref{lemma: scale} to construct the block-encoding of $\tau_k \Ibb$. Next, we use Lemma \ref{lemma: sumencoding} to construct the $\epsilon$-approximated block-encoding of $ \frac{1}{2}\left( \gamma + \tau_k \Ibb\right)$. Let $\kappa_{\tau_k}$ be the condition number of this operator. Then we use Lemma \ref{lemma: negative} with $c=1$ to obtain the $\mathcal{O}(\epsilon)$-approximated block-encoding of 
    $$\frac{(\gamma + \tau_k \Ibb)^{-1}}{ 2 \kappa_{\tau_k}}. $$ 
    As the next step, we use Lemma \ref{lemma: sumencoding} with appropriate coefficients to construct the $\mathcal{O}(\epsilon)$-approximated block-encoding of:
    \begin{align}
        \begin{split}
            &\frac{1}{\sum_{k=1}^m \omega_k  \kappa_{\tau_k} } \sum_{k=1}^m  \omega_k \kappa_{\tau_k} \frac{(\gamma + \tau_k \Ibb)^{-1}}{ 2 \kappa_{\tau_k}} \\
            &= \frac{1}{ 2\sum_{k=1}^m \omega_k  \kappa_{\tau_k}} \sum_{k=1}^m \omega_k (\gamma + \tau_k \Ibb)^{-1}
        \end{split}
    \end{align}
    Next, we use Lemma \ref{lemma: product} to construct the $\mathcal{O}(\epsilon)$-approximated block-encoding of 
    \begin{align}
        \frac{1}{ 2\sum_{k=1}^m \omega_k  \kappa_{\tau_k}}  \gamma\sum_{k=1}^m \omega_k (\gamma + \tau_k \Ibb)^{-1}
        \label{eqn: iii9}
    \end{align}
    As the last step, we use Lemma \ref{lemma: sumencoding} (with appropriate coefficients) combined with the block-encoding of $\gamma$ and the operator above, to construct the $\mathcal{O}(\epsilon)$-approximated block-encoding of:
    \begin{widetext}
         \begin{align}
         \label{iii10}
        \frac{1}{  \left( \sum_{k=1}^m \omega_k \alpha_k + 2\sum_{k=1}^m \omega_k  \kappa_{\tau_k} \right) } \left( (\sum_{k=1}^m \omega_k \alpha_k )\gamma  - \gamma \sum_{k=1}^m \omega_k (\gamma + \tau_k \Ibb)^{-1} \right)  
    \end{align}
    \end{widetext}
    which is exactly the $\mathcal{O}(\epsilon)$-approximated block-encoding of 
    \begin{align}
       \frac{1}{   \sum_{k=1}^m \omega_k \left(\alpha_k + 2   \kappa_{\tau_k} \right) }   \frac{ \gamma\log \gamma}{\log \kappa_{\gamma}}
       \label{iii11}
    \end{align}
    where $\kappa_{\gamma}$ is the condition number of $\gamma$. Then we can use Lemma \ref{lemma: amp_amp} to remove the factor $ \frac{1}{2}  \sum_{k=1}^m \omega_k \left(\alpha_k + 2   \kappa_{\tau_k} \right)$, resulting in the block-encoding of $ \gamma\log \gamma / 2\log (\kappa_\gamma)$.  
\end{itemize}


\vspace{2mm}
\noindent
\textbf{Step 4: Trace expectation.}  We point out that:
\begin{align}
\begin{split}
     D_{x\log x}(\rho\Vert\sigma)&=\text{Tr}[\sigma^{1/2} \, 4\kappa_{\sigma} \log( 4\kappa_{\sigma})\gamma\log \gamma \, \sigma^{1/2}]\\
&= 4\kappa_{\sigma} \log( 4\kappa_{\sigma})\text{Tr}\big[\sigma\, \gamma\log \gamma \big]. \\
\longrightarrow \frac{   D_{x\log x}(\rho\Vert\sigma)}{4\kappa_{\sigma} \log( 4\kappa_{\sigma})} &= \text{Tr}\big[\sigma\, \gamma\log \gamma \big].
\end{split}
\end{align}
Recall that in Lemma \ref{lemma: blockencodingdensityoperator}, there are two ways to block-encode the density operator of interest. As a consequence, there are two ways to estimate the above $f$-divergence. 
\begin{itemize}
    \item (Purification access Lemma \ref{lemma: blockencodingdensityoperator}) In this setting, we are given the unitary $U_{\sigma}$ that is the purification of $\sigma$, which can provide the block-encoding of $\frac{\pi \sigma}{4}$. Therefore, we can directly apply the method proposed in \cite{rall2020quantum}. Specifically, according to their Lemma 5, given the unitary $U_\sigma$ and block-encoding of $\gamma \log \gamma/2\log \kappa_\gamma$ the quantity 
   \begin{align}
         \begin{split}
        &\frac{\text{Tr}\big[\sigma\, \gamma\log \gamma \big]}{ 2 \log \kappa_{\gamma}} = \frac{  D_{x\log x}(\rho\Vert\sigma) }{  4\kappa_{\sigma} \log( 4\kappa_{\sigma}) 2 \log \kappa_{\gamma}} 
      \end{split}
      \label{iii13}
    \end{align}
    can be estimated to an additive accuracy $\epsilon$. The estimation of $D_{x\log x}(\rho\Vert\sigma) $ can be inferred from the above estimation by a multiplication of $ 4\kappa_{\sigma} \log( 4\kappa_{\sigma}) 2\log \kappa_{\gamma}$ via Route 2. 

    
    \item (Sample access \ref{lemma: blockencodingdensityoperator}) In this setting, we first need to use multiple copies of $\sigma$ to construct the (approximated) block-encoding of $\frac{\pi\sigma}{4}$. Then with the unitary block-encoding of $\sim \gamma \log \gamma$, we can use Lemma \ref{lemma: product} to construct the block-encoding of 
    \begin{align}
       \frac{ \pi \sigma \gamma\log \gamma}{4 \cdot 2\log \kappa_{\gamma}}. 
    \end{align}
    Then we can apply the method in \cite{subramanian2021quantum} to estimate the normalized trace
    \begin{align}
    \begin{split}
         &\frac{1}{N}    \frac{ \pi \Tr \left(\sigma \gamma\log \gamma \right) }{4 \log \kappa_{\gamma}} \\
        &=\frac{1}{N}   \frac{ \pi D_{x\log x}(\rho\Vert\sigma) }{16 \kappa_\sigma \log(4\kappa_\sigma) 2\log \kappa_{\gamma}}
    \end{split}
    \label{iii15}
    \end{align}
    to an additive error $\epsilon$ by repeating a quantum circuit of complexity $\mathcal{O}\left( T_\sigma + T_{\gamma \log \gamma} \right)$ a total of $\mathcal{O}(1/\epsilon^2)$ times. The estimation of the, namely, \textit{normalized $f$-divergence} $D_{x\log x}(\rho\Vert\sigma)/N $ can be inferred from the estimation above by an according multiplication. 

    We note that, in the same work \cite{subramanian2021quantum}, the authors also showed that from the estimation of $ D_{x\log x}(\rho\Vert\sigma)/N$ with additive precision $\epsilon$, such estimation with multiplicative precision $\delta $ can also be inferred. The details of the procedure can be found in Appendix B of \cite{subramanian2021quantum}.
\end{itemize}

\vspace{3mm}
 \begin{center}
    \textbf{Complexity analysis}
\end{center}
We recall $\gamma =\sigma^{-1/2}\rho\,\sigma^{-1/2}$ as defined earlier. For now, let the complexity for the block-encodings of $\rho,\sigma$ are $C_\rho,C_\sigma$. So the complexity for \textbf{Step 1} in Algorithm \ref{algo: algorithm1} is $C_\rho+C_\sigma $.  

Next, we analyze the complexity of \textbf{Step 2}. In this step, we first use Lemma \ref{lemma: negative} to construct the $\epsilon$-approximated block-encoding of $\sim\sigma^{-1/2}$, which has complexity $\mathcal{O}\left(  C_\sigma \kappa_\sigma \log \frac{1}{\epsilon} \right) $. Then we use Lemma \ref{lemma: product} to construct the $\epsilon$-approximated block-encoding of $\frac{1}{4\kappa_\sigma^2}\sigma^{-1/2} \rho \sigma^{-1/2} \equiv \gamma$. So, the circuit complexity for block-encoding $\gamma$ is $\mathcal{O}\left( C_\sigma \kappa_\sigma \log \frac{1}{\epsilon} + C_\rho \right) $. 

Next, in \textbf{Step 3}, we use the block-encoding from \textbf{Step 2} to construct the block-encoding of $ \sim\frac{\gamma \log \gamma}{ \log \kappa_\gamma}  $. We used two routes, based on two approaches in Lemma \ref{lemma: approximationoflog} to approximate $\log \gamma$. In Route 1, we use Lemma \ref{lemma: qsvt} with the polynomial transformation $p(.)$ having degree $\mathcal{O}\left(  \kappa_\gamma \log \frac{1}{\epsilon}  \right)$ to transform the block-encoded $\gamma$ into $p(\gamma)  \sim\gamma \log \gamma$. Thus, the circuit complexity via this route is $\mathcal{O}\left( \left( C_\sigma \kappa_\sigma \log \frac{1}{\epsilon} + C_\rho \right)  \kappa_\gamma \log \frac{1}{\epsilon} \right)$.

In Route 2, we first need to take the (approximated) block-encoding of $\gamma$ and use Lemma \ref{lemma: sumencoding} to construct the block-encoding of $ \frac{1}{2}(\gamma + \tau_k \Ibb)$. This step has complexity $\mathcal{O}\left( C_\sigma \kappa_\sigma \log \frac{1}{\epsilon} + C_\rho \right)$. Then we use the lemma \ref{lemma: negative} to transform the (block-encoded) $  \gamma + \tau_k \Ibb \longrightarrow \frac{(\gamma + \tau_k \Ibb)^{-1}}{ 2 \kappa_{\tau_k}}$, which incurs circuit complexity 
$$\mathcal{O}\left( \left( C_\sigma \kappa_\sigma \log \frac{1}{\epsilon} + C_\rho \right) \kappa_{\tau_k} \log \frac{1}{\epsilon} \right). $$
Then we use Lemma \ref{lemma: sumencoding} to construct the block-encoding of the summation in Eqn.~\ref{eqn: iii9}, followed by another application of the same lemma to obtain the block-encoding of $\sim \gamma \log \gamma$ (Eqn.~\ref{iii10}). There are totally $m$ terms in the summation, and hence requiring $m$ block-encodings of $\frac{(\gamma + \tau_k \Ibb)^{-1}}{ 2 \kappa_{\tau_k}} $. As indicated earlier, each block-encoding has complexity $\mathcal{O}\left( \left( C_\sigma \kappa_\sigma \log \frac{1}{\epsilon} + C_\rho \right) \kappa_{\tau_k} \log \frac{1}{\epsilon} \right) $, so the total circuit complexity at Eqn.~\ref{eqn: iii9} is:
\begin{align}
\begin{split}
    \mathcal{O}\left( \sum_{k=1}^m \left( C_\sigma \kappa_\sigma \log \frac{1}{\epsilon} + C_\rho \right) \kappa_{\tau_k} \log \frac{1}{\epsilon} \right) 
    \end{split}
\end{align}
We point out the following well-known property. Let $\lambda_{\max},\lambda_{\min}$ denotes the maximum  and minimum eigenvalues of $\gamma$ (in magnitude), then $\kappa_\gamma = \frac{\lambda_{\max}}{\lambda_{\min}}$. The condition number of $\frac{1}{2}(\gamma + \tau_k \Ibb) $ is:
\begin{align}
    \kappa_{\tau_k} = \frac{ \tau_k +\lambda_{\max}  }{\tau_k +\lambda_{\min} } <  \frac{\lambda_{\max}}{\lambda_{\min}}
\end{align}
which means that for all $k=1,2,...,m$, $ \kappa_{\tau_k}  \leq \kappa_\gamma $. So the complexity above can be reduced to:
\begin{align}
     \mathcal{O}\left(m  \left( C_\sigma \kappa_\sigma \log \frac{1}{\epsilon} + C_\rho \right) \kappa_\gamma \log \frac{1}{\epsilon}\right)
\end{align}

Recall from Lemma \ref{lemma: approximationoflog} that on the interval $(\frac{1}{\kappa_\gamma},1)$, the value of $m = \mathcal{O}\left(  \log(\kappa_\gamma) \log \frac{1}{\epsilon}  \right) $ is sufficient to guarantee the approximation error to be $\epsilon$, e.g., the operator in Eqn.~\ref{iii10} is $\epsilon$-close (in operator norm) to the one in Eqn.~\ref{iii11}. At the last step of Route 2, we use Lemma \ref{lemma: amp_amp} to remove the factor in the denominator of Eqn.~\ref{iii11}, so the total complexity then is:
\begin{align}
      \mathcal{O}\left( (   \sum_{k=1}^m \omega_k \left(\alpha_k + 2   \kappa_{\tau_k} \right))m  \left( C_\sigma \kappa_\sigma \log \frac{1}{\epsilon} + C_\rho \right) \kappa_\gamma \log \frac{1}{\epsilon}\right)
\end{align}
We note the following:  $\omega_k, \alpha_k \in \mathcal{O}(1)$ for all $k$, and hence $ \sum_{k=1}^m \omega_k\alpha_k \in \mathcal{O}(m)$; $ \sum_{k=1}^m \omega_k \kappa_{\tau_k} \in \mathcal{O}\left(\sum_{k=1}^m  \kappa_{\tau_k} \right) \in \mathcal{O}\left( m \kappa_\gamma \right)$; and $m = \mathcal{O}\left( \log ( \kappa_\gamma) \log \frac{1}{\epsilon} \right) $, so we have that the complexity above is:
\begin{align}
      &\mathcal{O}\left( m^2 \kappa^2_\gamma \left( C_\sigma \kappa_\sigma \log \frac{1}{\epsilon} + C_\rho \right)  \log \frac{1}{\epsilon}\right) \\
       &=\mathcal{O}\left(  \kappa^2_\gamma (\log \kappa_\gamma)^2 \left( C_\sigma \kappa_\sigma \log \frac{1}{\epsilon} + C_\rho \right)  \log^3 \frac{1}{\epsilon}\right) 
\end{align}
Therefore, as a summary, we have the following statements regarding the complexity to obtain the operator $ \gamma \log \gamma/(2\kappa_\gamma)$ within $\textbf{Step 3}$:
\begin{align}
\begin{split}
    \text{via Route 1: } T_{\gamma \log \gamma} = & \mathcal{O}\left(  \left( C_\sigma \kappa_\sigma \log \frac{1}{\epsilon} + C_\rho \right) \kappa_\gamma \log \frac{1}{\epsilon}\right)  \\
    \text{via Route 2: }T_{\gamma \log \gamma} =  & \mathcal{O}\left(  \left( C_\sigma \kappa_\sigma \log \frac{1}{\epsilon} + C_\rho \right) \kappa^2_\gamma \log^2(\kappa_\gamma)     \log^3 \frac{1}{\epsilon} \right) 
    \end{split}
    \label{iii20}
\end{align}

The final step, \textbf{Step 4}, is to estimate the desired quantity $D_{x\log x}(\rho\Vert\sigma) $. As outlined previously, the procedure and the corresponding complexity depend on the access to $\rho, \sigma$.
\begin{itemize}
    \item Via purification access: as indicated in \textbf{Step 4} of Algorithm \ref{algo: algorithm1}, we first use the algorithm in \cite{rall2020quantum} to estimate the quantity in Eqn.~\ref{iii13}. For an additive accuracy $\epsilon$, the algorithm has complexity $\mathcal{O}\left( (T_\sigma + T_{\gamma \log \gamma} ) \frac{1}{\epsilon    } \right)$
where $ T_\sigma, T_{\gamma \log \gamma} $ are the circuit complexity of $U_\sigma$ and block-encoding of $\sim \gamma \log \gamma$, respectively. Then we multiply this estimate by the factor in the denominator above to obtain the estimate of $ D_{x\log x}(\rho\Vert\sigma)  $. 

However, we note that the multiplication step will inflates the accuracy by the same factor. As a result, by a slight abuse of notation, if we target an $\epsilon$-additive accuracy of $ D_{x\log x}(\rho\Vert\sigma)$, we need to rescale the value of $\epsilon$ in the above estimation as
    $$ \epsilon \longrightarrow \frac{\epsilon}{4\kappa_{\sigma} \log( 4\kappa_{\sigma})  2\log \kappa_{\gamma}} $$
    Then the quantum circuit complexity is
    \begin{widetext}
\begin{align}
\begin{split}
     &\mathcal{O}\left( (T_\sigma + T_{\gamma \log \gamma} ) \frac{\kappa_{\sigma} \log( \kappa_{\sigma}) \log \kappa_{\gamma}}{\epsilon    } \right) 
     \end{split}
     \label{iii21}
\end{align}
    \end{widetext}

    In addition, we point out that in the purification access setting, we are provided with two unitaries $U_\sigma, U_\rho$ which are purification of $\sigma,\rho$ respectively. According to Lemma \ref{lemma: blockencodingdensityoperator}, the exact block-encoding of $\sigma,\rho$ can be constructed using $\mathcal{O}(1)$ $U_\sigma, U_\rho$ plus $\log N$ two-qubit gates. So, the circuit complexity of the block-encoding of $\sigma$ is: $C_\sigma = 2T_\sigma +\log N$. Likewise, $C_\rho= 2  T_\rho + \log N$ where we remind that $T_\sigma, T_\rho$ are the circuit complexity of $U_\sigma,U_\rho$. For convenience, we define $T \equiv \max \{  T_\sigma, T_\rho\}$. Replacing these complexities in Eqn.~\ref{iii20} and Eqn.~\ref{iii21}, we have that via Route 1 (direct QSVT), the circuit complexity for estimating quantum $f$-divergence $D_{x\log x}(\rho\Vert\sigma) $ is 
    $$  \mathcal{O} \left(   (2T + \log N)  \kappa^2_{\sigma} \kappa_\gamma\log( \kappa_{\sigma})  \log(\kappa_\gamma) \frac{1}{\epsilon}\log^2\frac{1}{\epsilon}\right).  $$
    Likewise, via Route 2 (Stieltjes representation), the circuit complexity: 
    $$  \mathcal{O} \left(   (2T + \log N)  \kappa^2_{\sigma} \kappa^2_\gamma\log( \kappa_{\sigma})  \log^3(\kappa_\gamma) \frac{1}{\epsilon}\log^4\frac{1}{\epsilon}\right).$$

    \item Via sample access: as indicated earlier, an application of the method in \cite{subramanian2021quantum} allows an estimation of quantity in Eqn.~\ref{iii15}. For a $\epsilon$-additive accuracy, the quantum circuit complexity is $\mathcal{O}\left( C_\sigma + T_{\gamma \log \gamma} \right) $ and the total number of repetition is $\mathcal{O}(1/\epsilon^2)$. From this estimate, the normalized $f$-divergence $\frac{1}{N}D_{x\log x}(\rho\Vert\sigma)$ can be inferred by a multiplication of $16 \kappa_\sigma \log(4\kappa_\sigma) \log \kappa_{\gamma}/\pi$. 
    
    We recall from Lemma \ref{lemma: blockencodingdensityoperator} that in the sample access setting, the circuit complexity for (approximately) block-encoding $\frac{\pi\sigma}{4}$ (and also $\frac{\pi\rho}{4}$) is $C_\sigma = C_\rho = \mathcal{O}\left(  \log (N) \frac{1}{\epsilon^2} \log \frac{1}{\epsilon}\right)   $. Therefore, via Route 1 (direct QSVT) of \textbf{Step 3}, the quantum circuit complexity is 
    $$\mathcal{O}\left(  \log (N)  \kappa_\sigma \kappa_\gamma \frac{1}{\epsilon^2} \log^3  \frac{1}{\epsilon}  \right) $$
At the same time, via Route 2, the quantum circuit complexity is 
$$ \mathcal{O}\left(  \log (N)  \kappa_\sigma \kappa^2_\gamma \log^2( \kappa_\gamma) \frac{1}{\epsilon^2} \log^5  \frac{1}{\epsilon}  \right)$$
    
    In both cases, the total of repetitions is the same, and it is:
 \begin{align}
 \begin{split}
       &\mathcal{O}\left( \frac{\kappa_{\sigma}^2 \log^2( 4\kappa_{\sigma})  \log^2 \kappa_{\gamma}}{\epsilon^2} \right)  
 \end{split}
    \end{align}
    
\end{itemize}

\subsection{General operator-convex $f$: Pad\'e-type (resolvent) approximations}
If $f$ is \emph{operator convex} on $(0,\infty)$ (with $f(1)=0$), the L\"owner--Kraus representation yields
\[
f(x)=a+bx+cx^2+\int_{0}^{\infty}\frac{(x-1)^2}{x+t}\,d\nu(t),
\]
for $a,b\in\mathbb{R}$, $c\ge0$, and a finite positive measure $\nu$.
Replacing the integral by a \emph{positive} $m$-point quadrature gives a Pad\'e-type rational
\[
r_m(x)=a+bx+cx^2+\sum_{k=1}^{m} \frac{w_k (x-1)^2}{x+t_k},\qquad w_k>0,\ t_k\ge0,
\]
which \emph{preserves operator convexity}. Hence for $A=\sigma^{-1/2}\rho\,\sigma^{-1/2}$,
\[
D_f(\rho\Vert\sigma)
= \text{Tr}\!\Big[\sigma^{1/2} f(A)\sigma^{1/2}\Big]
\;\approx\;
\text{Tr}\!\Big[\sigma^{1/2} r_m(A)\sigma^{1/2}\Big].
\]
and we have that:
\begin{align}
    r_m(A) = a\Ibb + b A + cA^2 + \sum_{k=1}^m \omega_k (A - \Ibb)^2 (A +\tau_k \Ibb)^{-1}.
\end{align}
Each term in $r_m(A)$ requires only matrix multiplications by $A$ and resolvents $(A+t_k I)^{-1}$, implementable with QSVT/LCU or quantum linear systems in a similar manner to what we describe above. Thus, \emph{any} operator-convex $f$ admits a structurally preserved, quantum-computable estimator via (i) block-encoding $A$, (ii) applying a rational approximation to $f$, and (iii) estimating the resulting trace expectation with amplitude estimation. The precision is governed by the rational approximation error for $f$ on $\mathrm{spec}(A)$; Positive quadrature weights ensure stability and preserve the relevant matrix inequalities. The complexity analysis essentially follows what we had above, except in \textbf{Step 3} we need to adjust the coefficients in the approximation of $f$ (via Route 2 -- Stieltjes representation). Therefore, the complexity is the same as what we obtained from the complexity of Route 2 above.

As indicated in Table \ref{tab: entropy}, by choosing appropriate function $f$ (and hence, the polynomial $p(.)$), the $f$-divergence reduces to the well-known entropies. Thus, our algorithm offers a universal framework for computing the quantum entropies, which were done separately before \cite{acharya2020estimating, wang2023quantum, nghiem2025estimation}.

\section{Quantum algorithms for Kubo-Ando means}
\label{sec: quantumalgorithm}

Building upon the progress discussed above, we extend our construction to the Kubo-Ando means. First, we give an introduction on this term, and provide a few examples.
\begin{definition}[Operator mean \cite{kubo1980means} ]

A binary operation $\sigma: B(H)^+ \times B(H)^+ \to B(H)^{++}$ is called an \emph{operator mean} in the sense of Kubo and Ando if it satisfies:

\begin{enumerate}
  \item \textbf{Monotonicity:} If $A_1 \le A_2$ and $B_1 \le B_2$, then $A_1 \,\sigma\, B_1 \le A_2 \,\sigma\, B_2$.
  \item \textbf{Transformer inequality:} For any operator $C$, $C (A \,\sigma\, B) C \le (C A C) \,\sigma\, (C B C)$.
  \item \textbf{Continuity:} If $A_n \downarrow A$ and $B_n \downarrow B$ in the strong operator topology, then $A_n \,\sigma\, B_n \downarrow A \,\sigma\, B$.
  \item \textbf{Normalization:} $I \,\sigma\, I = I$, where $I$ is the identity on $H$.
\end{enumerate}
Furthermore, there exists a unique operator-monotone function $f_\sigma: [0,\infty)\to[0,\infty)$ with $f_\sigma(1)=1$ such that, for invertible $A,B$,
\[
A \,\sigma\, B = A^{1/2}\, f_\sigma\bigl(A^{-1/2} B A^{-1/2}\bigr)\, A^{1/2}.
\]
\end{definition}
Examples include:
\begin{itemize}
    \item Arithmetic mean: \( A \nabla B = \frac{A + B}{2} \), \( f(x) = \frac{1 + x}{2} \).
\item Harmonic mean: \( A ! B = 2 (A^{-1} + B^{-1})^{-1} \), \( f(x) = \frac{2x}{1 + x} \).

\item  Geometric mean: \( A \# B = A^{1/2} (A^{-1/2} B A^{-1/2})^{1/2} A^{1/2} \), \( f(x) = \sqrt{x} \).

 \item Weighted spectral geometric mean: \( F_t(A, B) = (A^{-1} \sharp_t B)^{1/2} A^{2-2t} (A^{-1} \sharp_t B)^{1/2} \), where \( A^{-1} \sharp_t B = A^{-1/2} (A^{1/2} B A^{1/2})^t A^{-1/2} \), \( t \in [0,1] \) .
\end{itemize}

   \begin{widetext}
    
\begin{center}
\renewcommand{\arraystretch}{1.2}
\begin{tabular}{lllc}
\toprule
\label{table: exampleonoperatormean}
Name & Symbol & Representing function $f(x)$ & Operator formula \\

Arithmetic mean & $A\nabla_t B$ &
$(1-t)+t\,x$ &
$(1-t)A+tB$ \\
Harmonic mean & $A!_t B$ &
$\displaystyle \frac{x}{(1-t)x+t}$ &
$\big((1-t)A^{-1}+tB^{-1}\big)^{-1}$ \\
Geometric mean & $A\#_t B$ &
$x^{\,t}$ &
$A^{1/2} H^{\,t} A^{1/2}$ \\
Logarithmic mean & $A\,L\,B$ &
$\displaystyle \frac{x-1}{\log x}\ (f(1)=1)$ &
$A^{1/2}\, \frac{H- I}{\log H}\, A^{1/2}$ \\
Heinz mean & $\mathrm{Heinz}_t(A,B)$ &
$\displaystyle \tfrac12\big(x^{\,t}+x^{\,1-t}\big)$ &
$\tfrac12\big(A\#_t B + A\#_{1-t} B\big)$ \\
Power means & $P_{p,t}(A,B)$ &
$\big((1-t)+t\,x^{p}\big)^{\!1/p}$ ($p\neq0$) &
$A^{1/2}\big((1-t)I+t\,H^{p}\big)^{\!1/p}A^{1/2}$ \\
& (limits) &
$\exp\!\big(t\log x\big)$ ($p\to0$) &
$A\#_t B$ \\

\end{tabular}
\end{center}
  \end{widetext}

  As can be seen from the table above, the geometric mean is a special case of Kubo-Anbo mean, by choosing a proper representing function. Therefore, our work can be regarded as a generalization of the work \cite{liu2025quantum}.

\subsection{Kubo--Ando representation (harmonic–mean mixture)}
For an operator monotone $f$ on $(0,\infty)$ with $f(1)=1$, the Kubo--Ando mean
$\sigma_f$ admits a representing probability measure $\mu$ on $[0,1]$ such that
\begin{align}
\label{eq:KA-mixture}
\begin{split}
    A \,\sigma_f\, B &= \int_0^1 A!_{t} B\, d\mu(t), \\
A!_{t}B &= \big((1-t)A^{-1}+t\,B^{-1}\big)^{-1}.
\end{split}
\end{align}
Using the above property, the quantum algorithm for Kubo-Ando means, namely, harmonic-mean mixture approach, proceeds as follows. 

\begin{enumerate}
  \item \textbf{Block-encode $A^{-1}$ and $B^{-1}$ via QSVT.}
  Assume $(1,a,\varepsilon_A)$ and $(1,b,\varepsilon_B)$ block-encodings of $A,B\succ0$
  with spectra $\sigma(A), \sigma(B)$ in $[\delta,1]$ ($\delta\in(0,1]$). Apply QSVT with $g(x)=1/x$ on $[\delta,1]$
  to obtain block-encodings of $\delta A^{-1}$ and $\delta B^{-1}$; the (near-optimal) degree is
  \[
  d_{\mathrm{inv}}([\delta,1],\varepsilon) \;=\; \Theta\!\Big(\frac{1}{\delta}\,\log\frac{1}{\varepsilon}\Big).
  \]
  These encodings are re-used for all $t$’s in the quadrature below.
  
  \item \textbf{Form the convex combination $M_t=(1-t)A^{-1}+t\,B^{-1}$ by Lemma \ref{lemma: sumencoding}}
  Use Lemma \ref{lemma: sumencoding} with nonnegative weights $(1-t),t$
  producing a block-encoding of
  \[
    M_t \;=\; (1-t)\delta A^{-1} + t\delta \,B^{-1}.
  \]
  Since  $\sigma(A), \sigma(B) \in [\delta,1]$, it holds that $\sigma(A^{-1}),\sigma(B^{-1})\subset[1,\delta^{-1}]$, Weyl/convexity implies
  \[
    \sigma(M_t)\subset[\,1,\ \delta^{-1}\,] \quad \text{for all } t\in[0,1],
  \]
  which fixes the conditioning range for the next inversion.

  \item \textbf{Invert once by QSVT to obtain $A!_t B = M_t^{-1}$.}
  Apply QSVT, Lemma \ref{lemma: qsvt} with $g(x)=1/x$ on $[1,\delta^{-1}]$. The degree is
  \[
  d_{\mathrm{inv}}([1,\delta^{-1}],\varepsilon)\;=\;\Theta\!\Big(\frac{1}{\delta}\,\log\frac{1}{\varepsilon}\Big),
  \]
  giving a block-encoding of $\delta  M_t^{-1}=A!_t B =\big((1-t)A^{-1}+t\,B^{-1}\big)^{-1} $ with error $O(\varepsilon)$ (per $t$).
  \item To approximate 
\[
\int_0^1 A!_t B\, d\mu(t)
\]
we choose a quadrature $\{(t_j,w_j)\}_{j=1}^m$ with $\sum_j w_j=1$:
\[
\int_0^1 A!_t B\, d\mu(t)\approx \sum_{j=1}^m w_j\,A!_{t_j}B
\]
By combining the $m$ block-encodings $\{A!_{t_j}B\}$ with Lemma \ref{lemma: sumencoding} with weights $\{w_j\}$, we then form the block-encoding of:
\begin{align}
    \frac{1}{\sum_{j=1}^m \omega_j} \sum_{j=1}^m w_j\,A!_{t_j}B =  \sum_{j=1}^m w_j\,A!_{t_j}B
\end{align}
\end{enumerate}

\medskip
\noindent\textbf{Complexity analysis:} Let $C_A,C_B$ be the complexity of circuit that block-encodes $A$ and $B$. Then in Step 1, the complexity for obtaining the block-encoding of $\delta A^{-1}, \delta B^{-1}$ is 
$$ \mathcal{O}\left(   C_A \frac{1}{\delta} \log \frac{1}{\epsilon} \right) ,\mathcal{O}\left(   C_B \frac{1}{\delta} \log \frac{1}{\epsilon} \right) $$
At second step, we use Lemma \ref{lemma: sumencoding}, which uses the block-encoding of $\delta A^{-1}, \delta B^{-1}$ each once, so the total complexity is
$$\mathcal{O}\left(   (C_A + C_B) \frac{1}{\delta} \log \frac{1}{\epsilon} \right) $$
In Step 3, we use Lemma \ref{lemma: qsvt} to invert the (block-encoded) operator $M_t$, to obtain the block-encoding of operator $\big( (1-t) A^{-1} + t B^{-1}\big)^{-1}$, incurring a total complexity 
$$\mathcal{O}\left(   (C_A + C_B) \frac{1}{\delta^2} \log^2 \frac{1}{\epsilon} \right) $$
The final step is to form the summation of $m$ different block-encoded operators, each with complexity above, so the total complexity is
$$ \mathcal{O}\left(  m  (C_A + C_B) \frac{1}{\delta^2} \log^2 \frac{1}{\epsilon} \right)  $$


\subsection{Löwner--Stieltjes (resolvent) representation}
Every operator monotone $f$ on $(0,\infty)$ has a (normalized) Stieltjes form
\begin{equation}\label{eq:stieltjes}
f(x) \;=\; \alpha + \beta x \;+\; \int_{(0,\infty)} \frac{x}{x+\lambda}\, d\nu(\lambda),
\quad \alpha,\beta\ge 0,
\end{equation}
hence for $X\succ 0$,
\begin{equation}\label{eq:stieltjes-matrix}
f(X) \;=\; \alpha I + \beta X \;+\; \int_{(0,\infty)} X(X+\lambda I)^{-1}\, d\nu(\lambda).
\end{equation}
Applied to $X:=A^{-1/2}BA^{-1/2}$:
\begin{align} 
\begin{split}
    A \,\sigma_f\, B &= A^{1/2}\, f\!\big(A^{-1/2}BA^{-1/2}\big)\, A^{1/2} \\
&= \alpha A + \beta B + A^{1/2}\!\left(\int \frac{X}{X+\lambda I}\, d\nu(\lambda)\right)\!A^{1/2}.
\end{split}
\end{align}   
\medskip

 \begin{widetext}
     \[
f(x)=\alpha+\beta x+\int_{(0,\infty)}\frac{x}{x+\lambda}\,d\nu(\lambda)
\quad\Rightarrow\quad
A\sigma_f B
= A^{1/2}\Big(\alpha I+\beta X+\!\int \tfrac{X}{X+\lambda I}\,d\nu(\lambda)\Big)A^{1/2},
\]
 \end{widetext}
where $X:=A^{-1/2}BA^{-1/2}$.

Based on the above property, the quantum algorithm for Kubo-Ando means via Löwner--Stieltjes representation approach, proceeds as follows.

\begin{enumerate}
  \item \textbf{Block-encode $A^{\pm 1/2}$ and form $X=A^{-1/2}BA^{-1/2}$.}
  Assume $(1,a,\varepsilon_A)$ and $(1,b,\varepsilon_B)$ block-encodings of $A$ and $B$, with spectra in $[\delta,1]$.
  Using Lemma \ref{lemma: positive} and \ref{lemma: negative} with $c=\pm \frac{1}{2}$  to implement the block-encodings of $A^{1/2}$ and $\delta^{1/2} A^{-1/2}$. The quantum circuit complexity is: 
  \[
 \mathcal{O}\!\big( C_A \frac{1}{\delta}\log \frac{1}{\epsilon}\big).
  \]
  Use Lemma \ref{lemma: product} to get the block-encoding of
  \(
 \delta X = \delta A^{-1/2} B A^{-1/2}
  \)
  with the total circuit complexity 
  $$  \mathcal{O}\!\big( C_A \frac{1}{\delta}\log \frac{1}{\epsilon} +C_B\big).$$

  \item \textbf{Shifted inverse: $(X+\lambda I)^{-1}$ by QSVT.}
  Fix a quadrature $\{(\lambda_k,w_k)\}_{k=1}^m$ that approximates the measure $d\nu$ (see Lemma \ref{lem:rat-approx-operator-monotone}).
  For each $\lambda_k>0$, first use Lemma \ref{lemma: scale} to construct the block-encoding of $\delta \lambda_k \Ibb$. Then use Lemma \ref{lemma: sumencoding} to build a block-encoding of $\frac{1}{2} \delta\big( X+\lambda_k \Ibb\big)$. The circuit complexity of this block-encoding is
   $$  \mathcal{O}\!\big( C_A \frac{1}{\delta}\log \frac{1}{\epsilon} +C_B\big)$$
  
 We seek to approximate $x\mapsto 1/x$ on the interval
  \[
  \big[\lambda_k+\lambda_{\min}(X),\, \lambda_k+\|X\|\big]
  \supseteq
  \big[\lambda_k+\delta,\, \lambda_k+1/\delta\big]
  \]
  by a polynomial of degree
  \[
  d_{\mathrm{inv}}^{(k)} \;=\; \Theta\!\Big( \frac{\lambda_k+1/\delta}{\lambda_k+\delta}\,\log\frac{m}{\varepsilon}\Big),
  \]
  Then we use Lemma \ref{lemma: matrixinversion} to invert the block-encoding of $\frac{1}{2}\delta \big( X+\lambda_k \Ibb\big) $, yielding a block-encoding of $ \left( \frac{\lambda_k+\delta}{\lambda_k + 1/\delta} \right) \frac{1}{\delta} (X+\lambda_k I)^{-1}$ with error $O(\varepsilon/m)$. The complexity of this step is
  $$  \mathcal{O}\! \left( \big(  C_A \frac{1}{\delta}\log \frac{1}{\epsilon} +C_B\big) \frac{\lambda_k+1/\delta}{\lambda_k+\delta}\,\log\frac{m}{\varepsilon}  \right) $$

  \item \textbf{Form $X(X+\lambda_k I)^{-1}$.}
  Compose the block-encodings from Steps 1–2 via Lemma \ref{lemma: product} to obtain a block-encoding of
  $$ F_k:= \left( \frac{\lambda_k+\delta}{\lambda_k + 1/\delta} \right)X(X+\lambda_k I)^{-1} $$
  with the total circuit complexity
  $$ \mathcal{O}\! \left( \big(  C_A \frac{1}{\delta}\log \frac{1}{\epsilon} +C_B\big) \frac{\lambda_k+1/\delta}{\lambda_k+\delta}\,\log\frac{m}{\varepsilon}  \right) $$
  
  \item \textbf{Assemble $r_m(X)$.}
  Use Lemma \ref{lemma: sumencoding} with nonnegative weights $\{\alpha,\beta,w_k\}$  to combine $\{I,\,X,\,F_1,\dots,F_m\}$ into a block-encoding of
  \begin{align}
      \begin{split}
         &\frac{\alpha I + \beta X + \sum_{k=1}^m w_k\,X(X+\lambda_k I)^{-1} }{ \alpha + \beta  + \sum_{j=1}^m \omega_k \frac{\lambda_k+1/\delta}{\lambda_k+\delta} } \\
         &= \frac{r_m(X)}{  \alpha + \beta  + \sum_{j=1}^m \omega_k \frac{\lambda_k+1/\delta}{\lambda_k+\delta}}
      \end{split}
  \end{align}
  where $r_m$ is the quadrature approximation of $f$ on $[\delta,1]$. 
  
    Next, we use Lemma \ref{lemma: amp_amp} to remove the factor in the denominator above, resulting in the block-encoding of $r_m(X)$. This step uses the block-encoding from Step 3 totally $m$ times, plus the block-encoding of $X$ one time, so the total complexity is:
    \begin{align}
         \mathcal{O}\! \left( \sum_{k=1}^m \big(  C_A \frac{1}{\delta}\log \frac{1}{\epsilon} +C_B\big) \frac{\lambda_k+1/\delta}{\lambda_k+\delta}\,\log\frac{m}{\varepsilon}  \right) 
    \end{align}
    As we have pointed out earlier, the ratio 
    $$ \frac{\lambda_k+1/\delta}{\lambda_k+\delta} \leq \frac{1}{\delta^2}$$
    for all $k$. So the complexity above can be reduced to: 
       \begin{align}
         \mathcal{O}\! \left(  m \big(  C_A \frac{1}{\delta}\log \frac{1}{\epsilon} +C_B\big) \frac{1}{\delta^2}\,\log\frac{m}{\varepsilon}  \right) 
    \end{align}
    Because each of the operator $F_k$ in Step 3 has error $\mathcal{O}(\epsilon/m)$, the total accumulated error in this step is $\mathcal{O}(\epsilon)$. 
  \item \textbf{Dress back with $A^{1/2}$.}
  Finally, use Lemma \ref{lemma: product} with the two $A^{1/2}$ block-encodings from Step 1 and the one above to obtain a block-encoding of
  \[
  A^{1/2}\,r_m(X)\,A^{1/2} \;\approx\; A\sigma_f B,
  \]

\end{enumerate}

\noindent\textbf{Complexity summary.} The final step uses one block-encoding of $r_m(X)$, which has complexity 
$$ \mathcal{O}\! \left(  m \big(  C_A \frac{1}{\delta}\log \frac{1}{\epsilon} +C_B\big)\frac{1}{\delta^2}\,\log\frac{m}{\varepsilon}  \right)  $$
and two block-encodings of $A^{1/2}$, which has complexity $ \mathcal{O}\!\big( C_A \frac{1}{\delta}\log \frac{1}{\epsilon}\big)$. So the total complexity is
$$ \mathcal{O}\! \left(  m \big(  C_A \frac{1}{\delta}\log \frac{1}{\epsilon} +C_B\big)\frac{1}{\delta^2} \,\log\frac{m}{\varepsilon}  \right) $$
According to Lemma \ref{lem:rat-approx-operator-monotone}, the value of $m$ is $\mathcal{O}\left(  \log \frac{1}{\epsilon}\right) $ suffices to approximate $f(x)$ to an accuracy $\epsilon$, e.g., $| f(x) - r_m(x) | \leq\epsilon$, which implies that $| r_m(X) - f(X) | _{\rm operator} \leq \epsilon$. So, the final complexity is
$$ \mathcal{O}\! \left(  \big(  C_A \frac{1}{\delta}\log \frac{1}{\epsilon} +C_B\big)\frac{1}{\delta^2}\,\log^2 (\frac{1}{\varepsilon} \log\frac{1}{\epsilon})\right) $$

   \section{Conclusion}
\label{sec: conclusion}
In this work, we have constructed two quantum algorithms for computing quantum $f$-divergence and Kubo-Ando means. Our work is inspired by previous progress \cite{nghiem2025estimation, acharya2020estimating, subramanian2021quantum, wang2023quantum, liu2025quantum}, and is technically focused on the block-encoding framework \cite{gilyen2019quantum}. We specifically show that, provided a unitary purification of $\rho,\sigma$, or independent copies of $\rho,\sigma$, the block-encoding of these density matrices can be obtained. From then on, we could leverage many existing recipes within the block-encoding framework to perform the transformation on these block-encoded density operators. The desired $f$-divergence $D_f$ can finally be estimated by using amplitude estimation. The same line of technique can be applied to construct the Kubo-Ando means, given the block-encoded matrices $A,B$.

As emphasized earlier, the quantum $f$-divergence, as well as Kubo-Ando means, are general quantities, which includes many well-known entropies, and geometric mean as special cases. As such, our algorithm can be thought of as a generalization of existing works \cite{nghiem2025estimation, acharya2020estimating, subramanian2021quantum, wang2023quantum, liu2025quantum}. Altogether, these results and ours have marked further steps regarding quantum algorithm for computing physical quantities, thus expanding the scope of quantum computing application towards physical problems. In \cite{liu2025quantum}, the authors have discussed a few examples of geometric mean in the field of machine learning and propose the corresponding quantum algorithm. What kind of application that our Kubo-Ando means can be useful for is highly interesting and we leave it for future exploration.

\section*{Acknowledgement}
This work was supported by the U.S. Department of Energy, Office of Science, National Quantum Information Science Research Centers, Co-design Center for Quantum Advantage (C2QA) under Contract No. DE-SC0012704. N.A.N. also acknowledge support from the Center for Distributed Quantum Processing at Stony Brook University. N.A.N. thanks the hospitality of Harvard University where he has an academic visit during the completion of this project.

\appendix
\section{Preliminaries}
\label{sec: preliminaries}
We provide a more formal summary of key concepts, including block-encodings, QSVT, Kubo--Ando operator means, and Riemannian metrics on \( \mathbb{S}_{++}^n \).

\subsection{Block-Encodings  }
The block encoding method is one of the most powerful tools in quantum computing because it allows us to embed a general matrix into a larger unitary operator, which can then be manipulated using quantum algorithms. This framework is particularly important since quantum computers are designed to work with unitary operations, but most useful matrices (such as Hamiltonians, stochastic matrices, or data matrices) are not unitary. By encoding these matrices into blocks of a unitary, one can leverage techniques like Quantum Singular Value Transformation (QSVT) and Quantum Linear Algebra algorithms to perform tasks such as matrix inversion, eigenvalue estimation, and computing functions of matrices efficiently \cite{gilyen2019quantum, childs2017quantum}.

The strength of block encoding lies in its generality and modularity: once a matrix is block encoded, a wide range of quantum algorithms can act on it with optimal complexity guarantees. This makes it a unifying framework that connects linear algebra, optimization, and quantum machine learning. For instance, algorithms for solving systems of linear equations, principal component analysis, and quantum simulation can all be expressed within the block encoding paradigm. In short, block encoding is not just a technical trick—it is a foundational bridge that transforms mathematical objects into quantum operations, enabling scalable algorithms that surpass classical limits.

\begin{definition}[Block-encoding \cite{childs2017quantum, gilyen2019quantum}]
Let $A$ be an $N \times N$ matrix with $N = 2^n$. A unitary $U_A$ acting on $n+a$ qubits is called an \emph{$(\alpha,a,\varepsilon)$-block-encoding} (or shortly $\epsilon$-approximated block-encoding) of $A$ if
\[
\left\| A - \alpha \, \bigl(\langle 0|^{\otimes a} \otimes I\bigr) \, U_A \, \bigl(|0\rangle^{\otimes a} \otimes I\bigr) \right\|_2 \leq \varepsilon,
\]
where $\alpha > 0$ is a normalization factor, $a$ is the number of ancilla qubits, and $\varepsilon \geq 0$ is the approximation error. 

In the special case $\varepsilon = 0$, $U_A$ is called an \emph{$(\alpha,a)$-block-encoding} of $A$, or an exact block-encoding of $A$ as we used in the main text.

\begin{remark}[Properties of block-encoding unitary]
\label{remark: propertyofblockencoding}
The block-encoding framework has the following immediate consequences:
\begin{enumerate}[label=(\roman*)]
    \item Any unitary $U$ is trivially an exact block encoding of itself.
    \item If $U$ is a block encoding of $A$, then so is $\Ibb_m \otimes U$ for any $m \geq 1$.
    \item The identity matrix $\Ibb_m$ can be trivially block encoded, for example, by $\sigma_z \otimes \Ibb_m$.
\end{enumerate}
\end{remark}

\end{definition}

\subsection{Quantum singular value transformation}
Quantum Singular Value Transformation (QSVT) provides a unified framework for applying polynomial transformations to the singular values of a block-encoded matrix. It generalizes and unifies many quantum linear algebraic algorithms, including Hamiltonian simulation, matrix inversion, and low-rank approximation.

\begin{theorem}[Quantum Singular Value Transformation \cite{gilyen2019quantum}]
\label{lemma: qsvt}
Let $U$ be a unitary acting on $a+n$ qubits that is an $(\alpha, a, 0)$-block encoding of an $N \times N$ matrix $A$. Suppose $p$ is a polynomial of degree $d$ such that
\[
|p(x)| \leq 1 \quad \text{for all } x \in [-1,1].
\]
Then there exists a sequence of single-qubit controlled phase rotations $\Phi = (\phi_1,\ldots,\phi_d)$ such that a quantum circuit of length $O(d)$, using controlled applications of $U$ and $U^\dagger$, implements a unitary $U_{\Phi}$ that is an $(1, a+1, \varepsilon)$-block encoding of $p(A/\alpha)$, with $\varepsilon$ arbitrarily small.
\end{theorem}

Given block-encoding of a matrix, say, $A$, QSVT enables polynomial approximations of functions like \( A^t \), \( \sqrt{A} \), \( \log A \), and \( \exp A \), with degrees logarithmic in precision \( \varepsilon \), which has been used in the recent paper of \cite{liu2025quantum}.

On the interval $(0,\infty)$, it is well-known that any \emph{operator monotone} function \(f\) satisfying the normalization \(f(1)=1\) also obeys the bound
\[
|f(x)| \leq 1 \quad \text{for all } x \in (0,1),
\]
since such functions are concave and non-decreasing on \((0,\infty)\). This ensures that \(f\) can be uniformly approximated by bounded polynomials on \((0,1]\), making it compatible with the requirements of the QSVT theorem. Hence, one can construct an efficient QSVT circuit to implement any operator monotone function \(f(A)\) for a positive-definite, block-encoded matrix \(A\), with complexity governed by the polynomial approximation error.
 
As a summary, QSVT allows one to approximate $p(A)$ for any polynomial $p$ by a quantum circuit of size $O(d)$, where $d$ is the degree of the polynomial. The complexity of QSVT-based algorithms therefore reduces to the degree of the polynomial required to approximate the target function:
\begin{itemize}
    \item For smooth analytic functions (e.g., exponential), the degree is $O(\log(1/\varepsilon))$.
    \item For discontinuous functions (e.g., sign function), the degree is $O(1/\varepsilon)$.
\end{itemize}

\noindent

\subsection{Operations involving block-encoded operators}

For convenience of readers, we collect a few helpful lemmas below. 

\begin{lemma}[Product of block-encoded matrices [Lemma~30 of \cite{gilyen2019quantum}]
\label{lemma: product}
If $U$ is an $(\alpha, a, \delta)$-block-encoding of a matrix $A$ and $V$ is a $(\beta, b, \varepsilon)$-block-encoding of a matrix $B$, then there exists a unitary $W$ that is an $(\alpha\beta, a+b, \alpha\varepsilon+\beta\delta)$-block-encoding of $AB$. Moreover, $W$ can be implemented by one query to each of $U$ and $V$.
\end{lemma}

\begin{lemma}[Linear combination of block-encoded matrices \cite{gilyen2019quantum}]
\label{lemma: sumencoding}
Let $\alpha, \beta > 0$ be constants, and let $\vec{\gamma} \in \mathbb{R}^m$ be a vector such that $\|\vec{\gamma}\|_1 \leq 1$. Suppose that each $U_j$ is a $(1, a, \varepsilon)$-block-encoding of a matrix $A_j$ for $j=1,\dots,m$. Then there exists a unitary $U$ that is a $(\eta, a+O(\log m), \varepsilon)$-block-encoding of $\sum_{j=1}^m \gamma_j A_j$, where $\eta = \|\vec{\gamma}\|_1$. Moreover, $U$ can be implemented by one query to each $U_j$ and $O(m)$ additional gates.
\end{lemma}

The lemmas below are direct consequences of Lemma \ref{lemma: qsvt} with appropriate choice of function $p(x)$ (in Lemma \ref{lemma: qsvt}).
\begin{lemma}[Negative Power Exponent \cite{gilyen2019quantum}, \cite{chakraborty2018power}]
\label{lemma: negative}
    Given a block encoding of a positive matrix $\frac{\mathcal{M}}{\gamma}$ such that 
    $$ \frac{\Ibb}{\kappa_M} \leq \frac{\mathcal{M}}{\gamma}\leq \Ibb. $$
    then we can implement an $\epsilon$-approximated block encoding of $\mathcal{M}^{-c}/(2\kappa_M^c)$ in complexity $\mathcal{O}( \kappa_M T_M (1+c) \log(  \frac{\gamma }{\epsilon} ) )$ where $T_M$ is the complexity to obtain the block encoding of $\mathcal{M}$. 
\end{lemma}
In principle, the lemma above can be used to invert a block-encoded matrix, by choosing $c=1$. However, there exists a more efficient way to do so, by using another choice of function within Lemma \ref{lemma: qsvt}.  
\begin{lemma}[Matrix inversion, see e.g.~\cite{gilyen2019quantum, childs2017quantum}]
\label{lemma: matrixinversion}
Given a block encoding of some matrix $A$  with operator norm $||A|| \leq 1$ and block-encoding complexity $T_A$, then there is a quantum circuit producing an $\epsilon$-approximated block encoding of ${A^{-1}}/{\kappa}$ where $\kappa$ is the conditional number of $A$. The complexity of this quantum circuit is $\mathcal{O}\left( \kappa T_A \log \left({1}/{\epsilon}\right)\right)$. 
\end{lemma}

\begin{lemma}[Positive Power Exponent \cite{gilyen2019quantum},\cite{chakraborty2018power}]
\label{lemma: positive}
    Given a block encoding of a positive matrix $\mathcal{M}/\gamma$ such that 
    $$ \frac{\Ibb}{\kappa_M} \leq \frac{\mathcal{M}}{\gamma} \leq \Ibb. $$
   Let $c \in (0,1)$. Then we can implement an $\epsilon$-approximated block encoding of $(\mathcal{M}/\gamma)^c/2$ in time complexity $\mathcal{O}( \kappa_M T_M \log (\frac{ 1}{\epsilon})  )$, where $T_M$ is the complexity to obtain the block encoding of $\mathcal{M}/\gamma$. 
\end{lemma}
\begin{lemma}[Informal, Scaling multiplication of block-encoded operators] 
\label{lemma: scale}
    Given the block-encoding of some matrix $A$, the block-encoding of $A/p$ where $p > 1$ can be prepared with an extra $\mathcal{O}(1)$ cost.
\end{lemma}
\begin{lemma}[\cite{gilyen2019quantum} Theorem 30; \bf Amplification]\label{lemma: amp_amp}
Let $U$, $\Pi$, $\widetilde{\Pi} \in {\rm End}(\mathcal{H}_U)$ be linear operators on $\mathcal{H}_U$ such that $U$ is a unitary, and $\Pi$, $\widetilde{\Pi}$ are orthogonal projectors. 
Let $\gamma>1$ and $\delta,\epsilon \in (0,\frac{1}{2})$. 
Suppose that $\widetilde{\Pi}U\Pi=W \Sigma V^\dagger=\sum_{i}\varsigma_i\ket{w_i}\bra{v_i}$ is a singular value decomposition. 
Then there is an $m= \mathcal{O} \Big(\frac{\gamma}{\delta}
\log \left(\frac{\gamma}{\epsilon} \right)\Big)$ and an efficiently computable $\Phi\in\mathbb{R}^m$ such that
    \begin{align}
\left(\bra{+}\otimes\widetilde{\Pi}_{\leq\frac{1-\delta}{\gamma}}\right)U_\Phi \left(\ket{+}\otimes\Pi_{\leq\frac{1-\delta}{\gamma}}\right) \\ =\sum_{i\colon\varsigma_i\leq \frac{1-\delta}{\gamma} }\tilde{\varsigma}_i\ket{w_i}\bra{v_i} , \text{ where } \Big|\!\Big|\frac{\tilde{\varsigma}_i}{\gamma\varsigma_i}-1 \Big|\!\Big|\leq \epsilon.
\end{align}
Moreover, $U_\Phi$ can be implemented using a single ancilla qubit with $m$ uses of $U$ and $U^\dagger$, $m$ uses of C$_\Pi$NOT and $m$ uses of C$_{\widetilde{\Pi}}$NOT gates and $m$ single qubit gates.
Here,
\begin{itemize}
\item C$_\Pi$NOT$:=X \otimes \Pi + I \otimes (I - \Pi)$ and a similar definition for C$_{\widetilde{\Pi}}$NOT; see Definition 2 in \cite{gilyen2019quantum},
\item $U_\Phi$: alternating phase modulation sequence; see Definition 15 in \cite{gilyen2019quantum},
\item $\Pi_{\leq \delta}$, $\widetilde{\Pi}_{\leq \delta}$: singular value threshold projectors; see Definition 24 in \cite{gilyen2019quantum}.
\end{itemize}
\end{lemma}

  \section{Pade's approximation to operator monotone functions}

To extend to operator monotone functions, we need some approximation to it.

\begin{lemma}[Löwner (Pick–Nevanlinna) representation on $(0,\infty)$  (Thm.~V.4.5] of \cite{donoghue2012monotone}), (Eq.~(5.62) of \cite{bhatia2006riemannian, bhatia2009positive}),\cite{donoghue2012monotone, bhatia2019bures, bhatia2006riemannian, hansen2013fast} ]
A function $f:(0,\infty)\to\mathbb{R}$ is operator monotone if and only if it admits the integral representation
\[
f(x)=\alpha+\beta x+\int_{0}^{\infty}\frac{x}{x+\lambda}\,d\mu(\lambda),
\]
for some $\alpha\in\mathbb{R}$, $\beta\ge 0$, and a finite positive measure $\mu$ on $[0,\infty)$ for which the integral converges.
\end{lemma}

This is a standard consequence of approximating a Stieltjes transform via positive quadrature and Padé approximants for Stieltjes (Markov) functions. In particular:

\begin{theorem}[Uniform convergence of Padé approximants for Stieltjes-type functions; see { Theorem 1 of \cite{lopez1989convergence}}]
Let $f(z)=\int_0^\infty \frac{d\mu(t)}{z-t}$ be a Stieltjes transform of a positive measure $\mu$ with finitely many poles. Then the diagonal Padé approximants $[n/n]_f(z)$ converge uniformly to $f(z)$ on compact subsets of the complex plane away from the support of~$\mu$.
\end{theorem}
This establishes that rational approximations---e.g., those constructed via 
quadrature rules or discrete sampling of the integral---can uniformly approximate 
Stieltjes functions away from singularities. It underpins our use of positive rational 
quadrature/Padé schemes to approximate an operator-monotone function $f(x)$ 
(which extends analytically to a Stieltjes transform) on a compact interval $[\delta,1]$. 
Such approximations yield the kernel sum
\[
r_m(x) \;=\; \alpha + \beta x + \sum_{k=1}^m w_k \, \frac{x}{x+\lambda_k},
\]
with exponentially small uniform error in $m$.

\begin{lemma}[Positive rational approximation of operator-monotone functions]
\label{lem:rat-approx-operator-monotone}
Let $f$ be an operator-monotone function on $(0,\infty)$ which extends analytically 
to a Stieltjes transform. Then for any $\delta \in (0,1)$ and any $\varepsilon \in (0,1)$, 
there exists a positive rational approximation $r_m(x)$ of the form
\[
r_m(x) \;=\; \alpha + \beta x + \sum_{k=1}^m w_k \,\frac{x}{x+\lambda_k},
\]
with $\alpha,\beta \geq 0$, $w_k > 0$, and $\lambda_k > 0$, such that
\[
\sup_{x \in [\delta,1]} \, \big| f(x) - r_m(x) \big| \;\leq\; \varepsilon.
\]
Moreover, $m$ can be chosen $m=O\!\big(\log(1/\varepsilon)\big)$ for fixed $\delta$, and the nodes/weights can be computed by quadrature schemes for Stieltjes transforms (e.g., Gauss-type or multipoint Padé), which enjoy exponential convergence on compact $[\delta,1]$ away from the branch cut. 
\end{lemma}

\bibliography{ref.bib}
\bibliographystyle{unsrt}
\end{document}